\documentclass[modern]{aastex62}

\submitjournal{ApJ}

\shorttitle{Influence of the GD effect on the spectrum of NS}
\shortauthors{Majczyna et al.}

\begin{document}

\title{Influence of the gravitational darkening effect on the spectrum of a hot, rapidly 
rotating neutron star}

\correspondingauthor{Agnieszka Majczyna}
\email{agnieszka.majczyna@ncbj.gov.pl}

\author[0000-0003-0864-8779]{Agnieszka Majczyna}
\affil{National Centre for Nuclear Research, ul. Andrzeja So\l{}tana 7, 05-400 Otwock, Poland;}
\nocollaboration

\author{Jerzy Madej}
\affil{Astronomical Observatory, University of Warsaw, Al. Ujazdowskie 4, 00-478 Warszawa, Poland;}
\nocollaboration

\author{Agata R\'o\.za\'nska}
\affil{Nicolaus Copernicus Astronomical Centre, Polish Academy of Sciences, ul. Bartycka
18, 00-716 Warszawa, Poland;}
\nocollaboration

\author{Miros\l{}aw Nale\.zyty}
\affil{Astronomical Observatory, University of Warsaw, Al. Ujazdowskie 4, 00-478 Warszawa, Poland;}
\nocollaboration

\begin{abstract}
 
In this paper, we discuss the influence of the gravitational darkening effect on the emergent 
spectrum of a fast-rotating, flattened neutron star. Model atmosphere codes always calculate spectra 
of emergent intensities and fluxes emitted from the unit surface on the star in plane-parallel 
geometry. Here we took a step beyond that and calculated a small sample grid of theoretical spectra 
integrated over the distorted surface of a sample rotating neutron star seen by a distant observer 
at various inclination angles. We assumed parameters like two dimensionless angular velocities 
$\bar{\Omega}^2=0.30$ and 0.60, the effective temperature of a nonrotating star $T_{\rm 
eff}=2.20\times 10^7\,$K, the logarithm of the surface gravity of a spherical star $\log(g)=14.40$ 
(cgs), and inclination angles from $i=0^\circ$ to $i=90^\circ$ with step $\Delta i=10^\circ$. We 
assumed that the atmosphere consists of a mixture of hydrogen and helium with $M_{\rm H}=0.70$ and
$M_{\rm He}=0.30$. At each point on the neutron star surface, we calculated true intensities for 
local values of parameters ($T_{\rm eff}$ and $\log(g)$), and these monochromatic intensities are 
next integrated over the whole surface to obtain the emergent spectrum. In this paper, we compute 
for the first time theoretical spectra of the fast-rotating neutron star. Our work clearly shows 
that the gravitational darkening effect strongly influences the spectrum and should be included in 
realistic models of the atmospheres of rotating neutron stars.

\end{abstract}


\section{Introduction}

A star distorted by either the centrifugal force caused by rotation or by tidal forces from a 
secondary star in a close binary system radiates away radiative energy in a nonsymmetric way and 
exhibits variations of the radiation power over its surface. Consequently, the effective temperature 
on the stellar surface is not uniform. The effect was recognized by Hugo von Zeipel in 1924 
\citep{vonzeipel24}. He investigated the radiative stellar interior in hydrostatic equilibrium. The 
main conclusion of his work was that the locally emergent bolometric flux of the radiation of a 
distorted star is proportional to the local gravity. Assuming of Stefan-Boltzmann law, it could be 
written as $T_{\rm eff}\sim g^{1/4}$, known as the {\it von Zeipel law}. The results of von Zeipel 
are valid only for stars with fully radiative envelopes and hydrostatic equilibrium. This law must 
be written in the more general form as $T_{\rm eff}\propto g_{\rm eff}^\beta$, where $\beta$ is the 
{\it gravitational darkening exponent} (GDE). If the envelope of the star is not fully radiative, 
the gravitational darkening exponent differs from 1/4.

The literature on the theory of stellar atmospheres involves the following terms: the gravitational 
darkening and the limb darkening. We direct the attention of the reader to the basic difference 
between these two terms. Limb darkening is an optical effect seen both in spherical and distorted 
stars where the central part of the star's disk is brighter than the star's limb. This effect 
denotes that the specific intensity of the radiation emitted at some point on the surface is the 
largest in the vertical direction and decreases in directions approaching the horizontal plane for a 
single value of the effective temperature at this point. Gravitational darkening denotes that the 
value of the effective temperature (total luminosity) varies over the surface of the distorted star. 
The distortion of the star can be caused either by centrifugal forces due to rotation or by tidal 
interactions between two stars in a close binary system.

Gravitational darkening effects were studied for different aspects and for different types of 
objects \citep[see e.g.][]{djurasevic03,djurasevic06,desouza06}. Theoretical models of gravitational 
darkening for different types of objects were also developed.\cite{lucy67} investigated a fully 
convective envelope and found that $\beta\sim 0.08$. This is a very weak dependence of the 
effective temperature on the effective gravity. The results of \citet{lucy67}
research were confirmed by \cite{webbink76}, who used a simple opacity law. The author presented 
an approximation, which under the assumption of constant entropy of the envelope, gave results 
consistent with those obtained by \citet{lucy67}. Note, that all these above results were obtained
under assumptions of rigid and small rate of rotation. In the case of neutron stars, the latter 
assumption is not strictly valid.

Recently \cite{claret15} investigated gravitational darkening for neutron stars with various 
rotational models assumed. The author assumed a spherical shape of the neutron star and explored the 
dependence of $\beta$ on the temperature of the star. There is a weak dependence of the 
gravitational darkening exponent on the temperature below 6$\times 10^7$ K in all assumed models of 
the rotation. In the above paper, the author was concerned about the calculation of the 
gravitational darkening exponent and its dependence on the rotation law. The author did not 
calculate any model spectra of a rotating star, either normal or neutron star.

The equilibrium configurations of a slowly rotating neutron star for different central densities
$\rho_{\rm c}$ and circular angular velocities $\Omega$ were investigated by \cite{belvedere14}. The 
most interesting result of this paper was presented in their Figure 16, where the authors have shown 
the dependence of the eccentricity on the frequency of the neutron star rotation. What is important, 
for frequency typical for neutron stars in X-ray bursters the eccentricity is of the order of 0.6, 
which means that the ratio of the polar radius to the equatorial radius $R_{\rm p}/R_{\rm 
eq}\sim 0.8$. We expect that such flattening should give rise to  visible changes in the emerging 
spectrum.

The effective surface acceleration of the rotating neutron star was calculated by \cite{algendy14} 
by using useful approximate formulae. On the surface of an oblate star, the effective gravity 
acceleration is decreased by the centrifugal force at the equator and increased by the centrifugal 
flattening of the star. The authors have shown useful formulae on 
the effective gravity for the approxi mation of fast and slow rotation. These formulae weakly depend 
on the chosen equation of state of the neutron star matter. In the slow rotation approximation, the 
effective gravity depends only on the compactness ratio $M/R_{\rm eq}$, the dimensionless square of 
the angular velocity $\Omega^2 R_{\rm eq}^3/GM$, and the latitude on the surface of the star. For 
spin frequency $\sim$600~Hz difference of the effective gravity between the pole and equator is 
about 20\%. This effect is important not only for the spectrum calculation but, also for it 
influence on the Eddington luminosity of a neutron star. \cite{algendy14} have shown that for the 
assumed $\bar{\Omega^2}=0.1$, the expected difference between the Eddington luminosity at the pole 
and at the equator is about 10\%. This is crucial because X-ray bursts at the Eddington limits (PRE 
- photospheric radius expansion bursts) are used as the standard candles \citep{galloway08}.

\cite{suleimanov_ea20} investigated the spectra of slowly rotating neutron stars for a few models of 
the emission, from the simplest blackbody to their theoretical model atmospheres. Using their 
models, the authors calculated the dilution factor $w$ and the spectral hardening factor $f_c$ to 
obtain local specific intensities, which are next integrated over the surface of the star. Different 
rotational frequencies and inclination angles were investigated. The shape of the star and the 
local gravity are calculated under the approximation of slow rotation. General relativistic effects 
were also included.

We note here that \cite{jaroszynski86} modeled the external gravitational field and surface shape of 
neutron stars rotating with the single angular velocity $\Omega=0.27 GM/R^3$. \cite{jaroszynski86} 
investigated the influence of the rotation on the observed luminosity and spectrum of the rotating 
neutron star. For the assumed angular velocity, eccentricity, and other basic parameters of the 
star, spectra have been calculated. Even if the blackbody model was used for the spectrum, the light 
bending was included. The ray-tracing procedure was included by the author. The main conclusion of 
\citet{jaroszynski86} was that a two-temperature blackbody shape is needed to fully describe the 
spectrum of the rotating star. The author also showed that a small, 10\%, changing eccentricity 
cannot be seen in the emerging spectrum.

In this paper, we show how fast rotation and therefore the gravitational darkening effect influences 
the shape of the spectrum of the oblate neutron star. Emergent spectra of the rotating neutron star 
seen by a distant observer are calculated by the ATM24 code, which solves the full radiative 
transfer including Compton scattering \citep[see e.g.][]{majczyna05b, madej17}. We calculated models 
with different parameters like an eccentricity, angular velocities, or inclination angle. We show 
that these spectra differ from each other and from the spectra of the spherical star. Therefore, the 
gravitational darkening effect should be included in realistic models of the neutron star 
atmospheres. 

\section{Model atmospheres}
\subsection{Gravitational darkening}

To calculate model spectra including the gravitational darkening effect, we need to define the 
geometry of the problem, calculate the effective gravity over the surface of the oblate star, and 
the intensity spectra at each point on the star. We used the spherical coordinate system where 
$\theta=0^\circ$ denotes the latitude of the equator.

In the first approximation, we assume that the shape of the rotating neutron star could be described 
as an ellipsoid. The radius of this solid is
\begin{equation}
R(\theta)=R_{\rm eq}\sqrt{1-e^2\cos(90^{\circ}-\theta)},
\end{equation}
where $\theta$ is the latitude in spherical coordinates, $R_{\rm eq}$ is the 
equatorial radius, whereas $R_{\rm p}$ is the polar radius, and $e$ 
is eccentricity, defined as $e=\sqrt{(R_{\rm eq}^2-R_{\rm p}^2)/R_{\rm eq}^2}$.

The effective gravity on the surface of the oblate star depends on the latitude angle. To describe 
this dependence, we use the approximate formula given by \cite{algendy14}, Eq.~50 in their work, in 
the regime of the fast rotation. In the original equation, the colatitude angle was used, whereas 
in the present paper, we use latitude angle $\theta$, therefore, we  
replace $\theta^\prime=90^\circ-\theta$

\begin{eqnarray}
g(\theta)/g_{\rm 0}&=&1+(c_{\rm e}\bar{\Omega^2}+d_{\rm e}\bar{\Omega^4}+f_{\rm 
e}\bar{\Omega^6})\sin^2(90^\circ-\theta) 
\nonumber \\ 
&+&(c_{\rm p}\bar{\Omega^2}+d_{\rm p}\bar{\Omega^4}+f_{\rm p}\bar{\Omega^6}-d_{\rm 60}\bar{\Omega^4}
)\cos^2(90^\circ-\theta) 
+d_{\rm 60}\bar{\Omega^4}\cos(90^\circ-\theta),
\label{eq:geff}
\label{eq:geff}
\end{eqnarray}
where $g_{\rm 0}$ is the gravity of a nonrotating star, and the following 
coefficients with $x=M/R_{\rm eq}$ are
\begin{eqnarray}
c_{\rm e}&=&-0.791+0.776x, \nonumber \\
d_{\rm e}&=&-1.315x+2.431x^2, \nonumber \\
f_{\rm e}&=&-1.172x, \nonumber \\
c_{\rm p}&=&1.138-1.431x, \nonumber \\
d_{\rm p}&=&0.653x-2.864x^2, \nonumber \\
f_{\rm p}&=&0.975x, \nonumber \\
d_{60}&=&13.67x-27.13x^2. \nonumber \\
\end{eqnarray}

The dimensionless angular velocity $\bar{\Omega}$ is defined as
\begin{equation}
\bar{\Omega}=\Omega\left(\frac{R_{\rm eq}^3}{GM}\right)^{1/2}.
\end{equation}

\begin{figure}[!h]
\begin{center}
\includegraphics[scale=0.4]{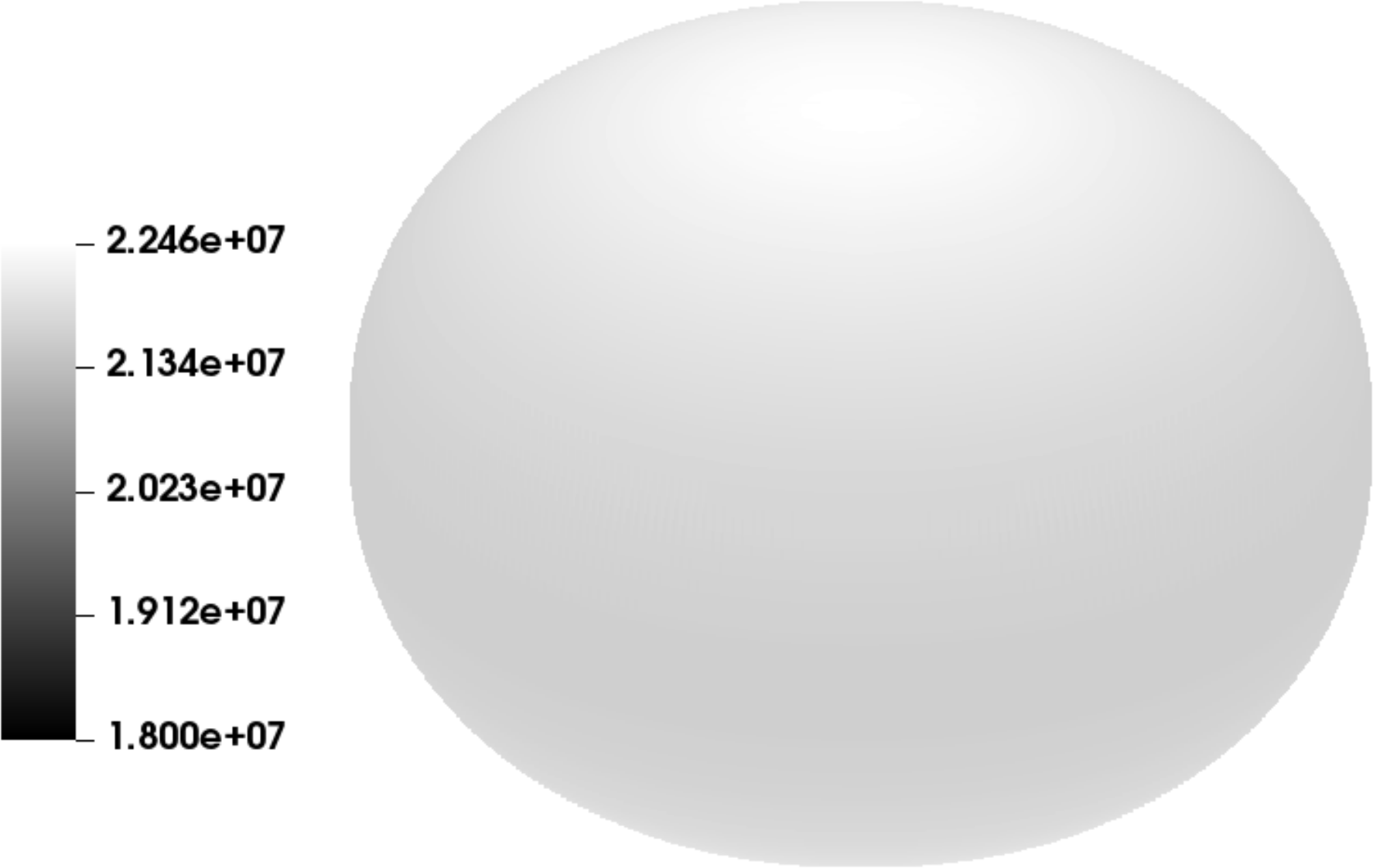}
\includegraphics[scale=0.4]{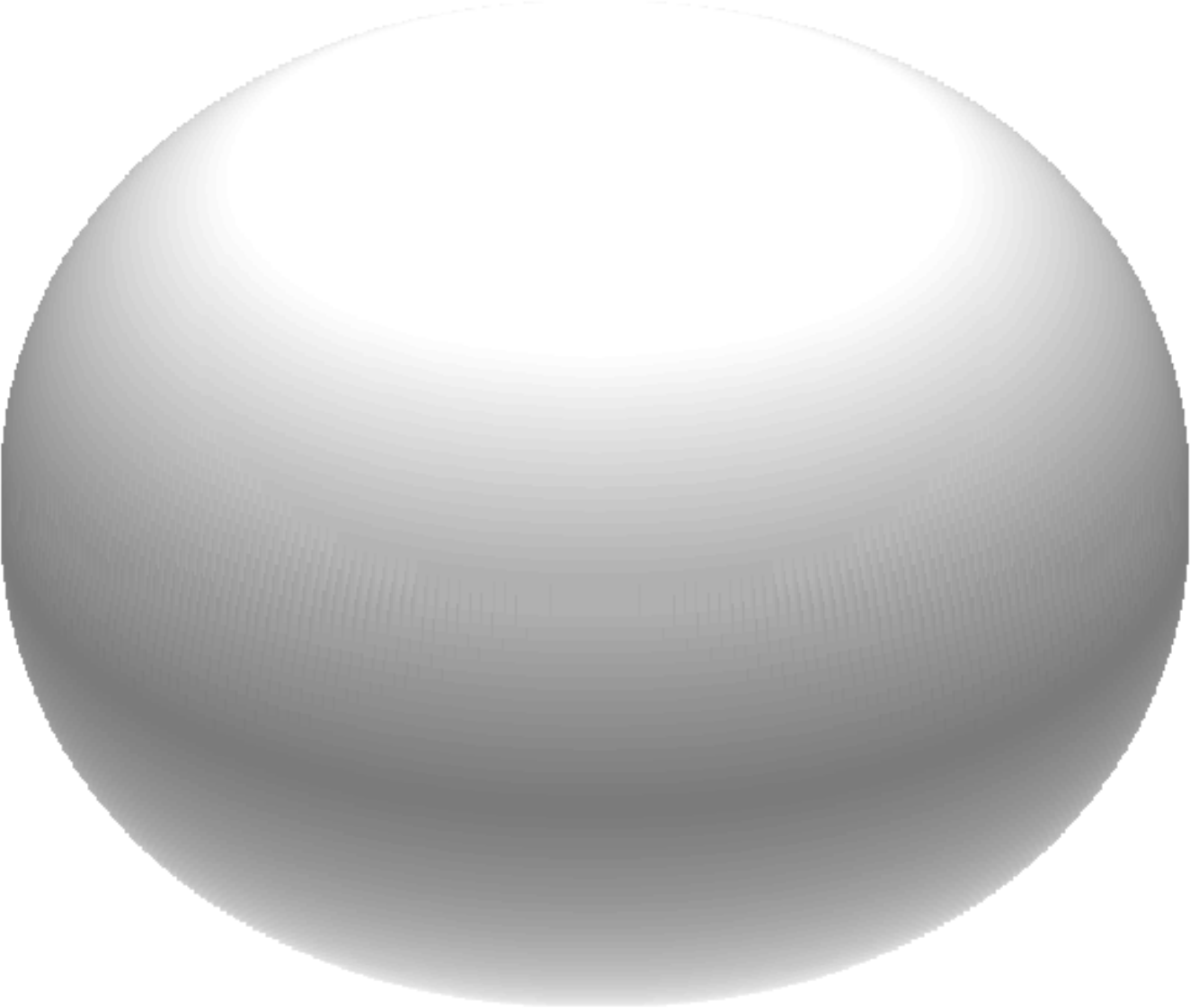}
\includegraphics[scale=0.4]{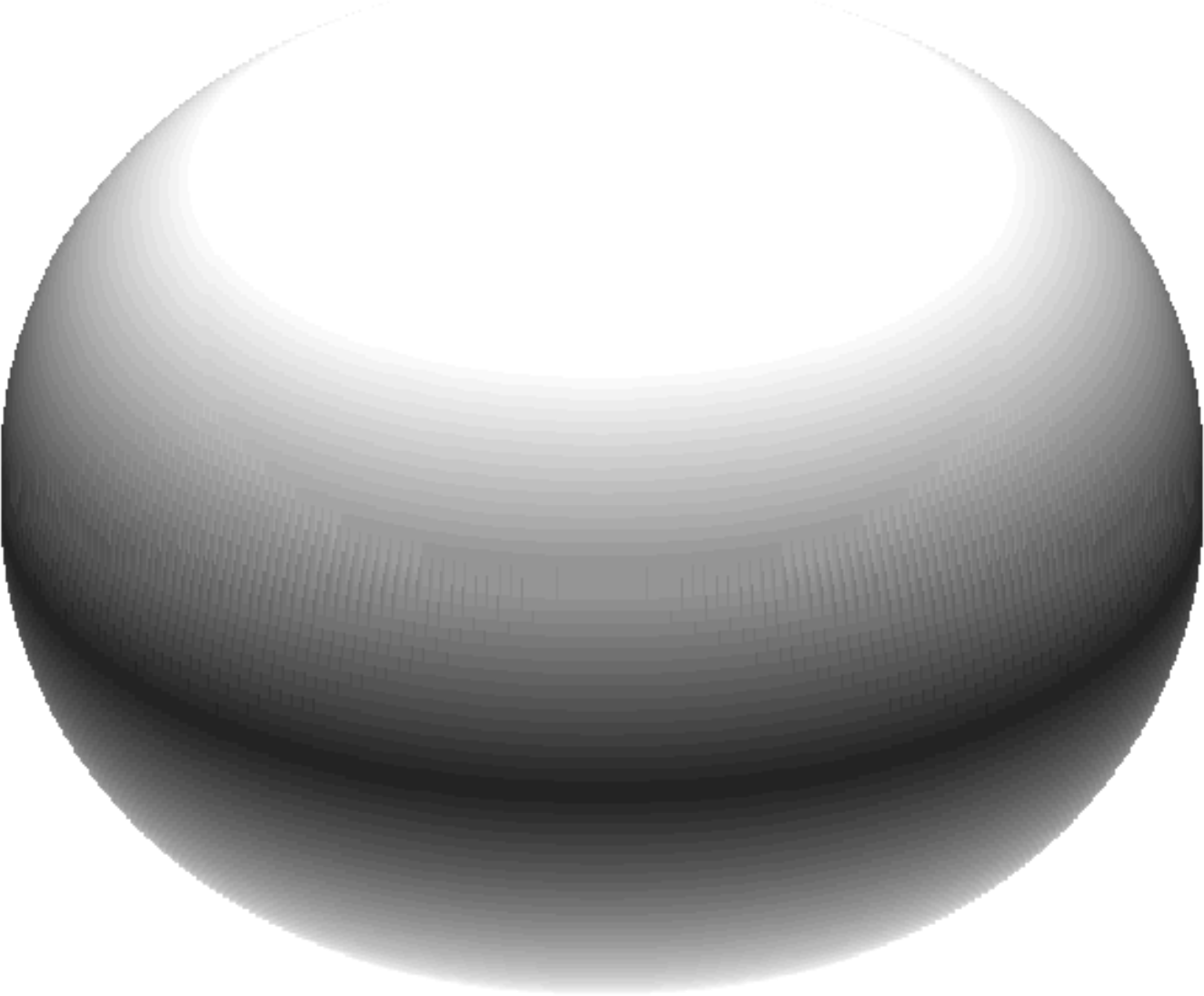}
\end{center}
\caption{Distribution of the effective temperature over the surface of a neutron star for 
different dimensionless angular velocities: $\bar{\Omega}^2=0.10$ (left panel),
$\bar{\Omega}^2=0.40$ (middle panel), and $\bar{\Omega}^2=0.60$ (right panel). 
We assumed the effective temperature and surface gravity of an undistorted star: 
$T_{\rm eff}=2.20\times
10^7\,$K and $\log(g)=14.40\,$(cgs), respectively, and Gravitational 
Darkening Exponent (GDE) $\beta=0.25$, the
eccentricity of the
rotating star $e=0.70$ and inclination angle $i=30^\circ$. }
\label{elipsoida-teff}
\end{figure}

The local effective gravity and therefore the local effective temperature are 
strong functions of the dimensionless angular velocity, which we have shown in 
the example of the distribution of the local temperature in Fig. 
\ref{elipsoida-teff}. In the figure, we use the same scale factor
for converting values of the temperature into the gray scale. For low values of 
the $\bar{\Omega}$, the local effective gravity or temperature varies with the 
latitude angle $\theta$ much slowly than
for higher values of $\bar{\Omega}$.

\begin{figure}[!h]
\begin{center}
\includegraphics[scale=0.38]{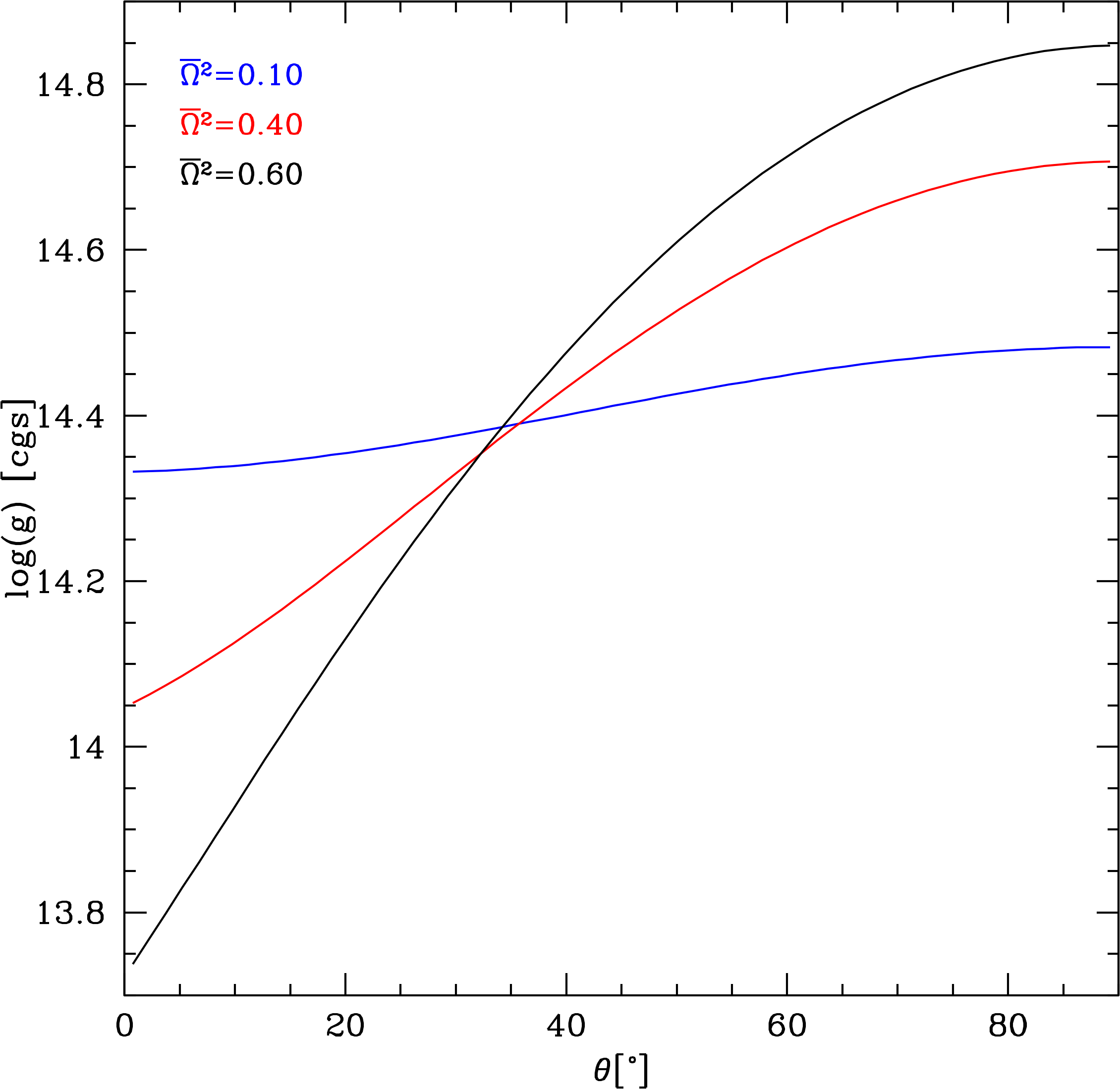}
\includegraphics[scale=0.38]{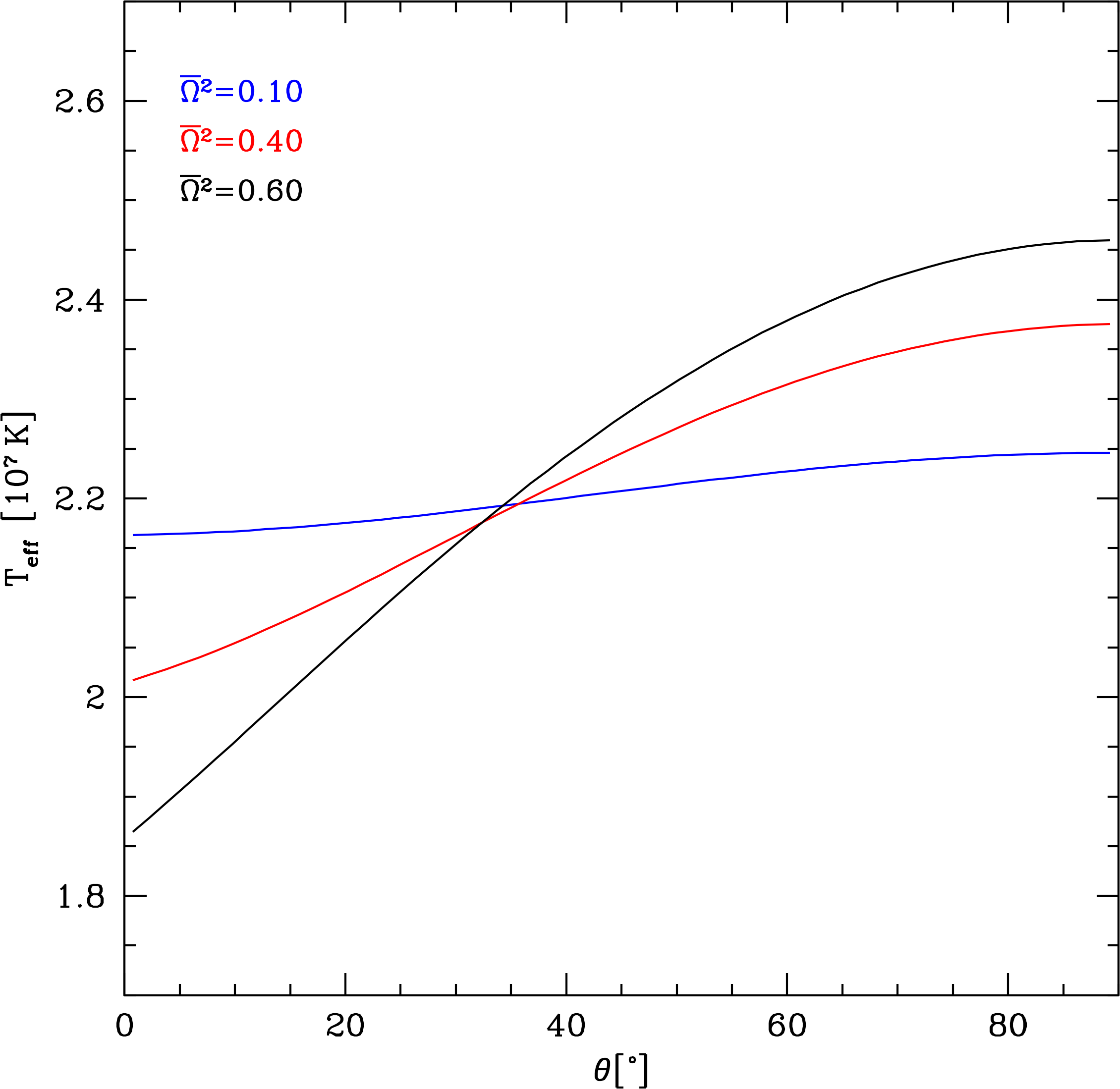}
\end{center}
\caption{Dependence of the effective gravity (left panel) and the effective 
temperature (right 
panel) on the latitude angle $\theta$ for various dimensionless angular 
velocities. We assume the logarithm of the surface gravity of a spherical star 
$\log(g)=14.40$(cgs), the effective 
temperature of the spherical star $T_{\rm eff}=2.20\times 10^7\,$K, GDE $\beta=0.25$ and various 
dimensionless angular velocities $\bar{\Omega}^2=0.10$ (blue line), 0.40 (red 
line), and 0.60 (black 
line).}
\label{geff-diffomega}
\end{figure}

Figure \ref{geff-diffomega} shows the dependence of the effective surface 
gravity on the latitude 
angle. We assume the gravity of a spherical star $\log(g)=14.40\,$(cgs). As was shown in the Fig.
\ref{geff-diffomega}, the surface gravity on the pole is not equal to the 
assumed surface gravity 
of a spherical star $\log(g)=14.40$. In our calculations, we include the 
quadrupole moment caused by 
the mass distribution in the neutron star interior. This moment does not 
vanish on the pole and causes 
greater gravity than the gravity of a nonrotating neutron star. 

\begin{figure}[!h]
\begin{center}
\includegraphics[scale=0.45]{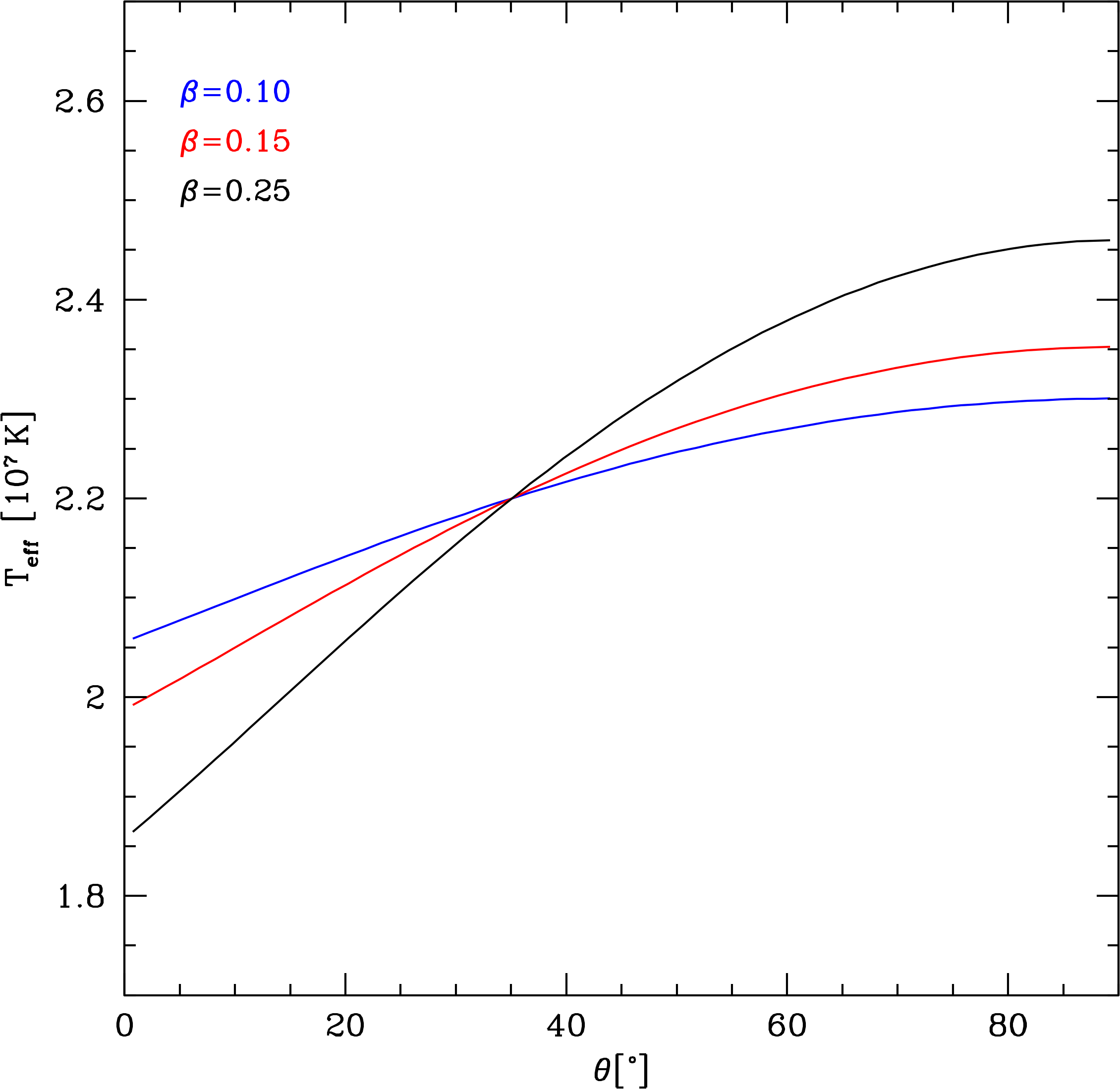}
\end{center}
\caption{Dependence of the effective temperature on the latitude angle 
$\theta$ for various 
GDEs: $\beta=0.25$ denoted by the black line, $\beta=0.15$ by the red one, and
$\beta=0.10$ by the blue one. We assume the effective temperature of 
a nonrotating, spherical star,
$T_{\rm eff}=2.20\times 10^7\,K$, and the logarithm of the surface gravity 
$\log(g)=14.40$(cgs) and
dimensionless angular velocity $\bar{\Omega}^2=0.60$.}
\label{teff}
\end{figure}

The temperature at a given point on the surface of the neutron star is as follows:
\begin{equation}
T(\theta)=T_{\rm eff}\left(\frac{g(\theta)}{g_0}\right)^{1/4},
\label{eq:teff}
\end{equation}
where $T_{\rm eff}$ is the effective temperature of the spherical, nonrotating 
neutron star, 
$g_0$ is the surface gravity of the spherical star, and $g(\theta)$ is the surface gravity at a 
given point of the distorted star defined by Eq.\ref{eq:geff}.

In the above Eq.\ref{eq:teff}, GDE is equal to 0.25 because we assumed that the 
atmosphere is 
radiative. As was shown by \citep[e.g.,][]{lucy67}, this exponent is different 
for nonradiative or 
not fully radiative envelopes. Very recently \cite{claret21} calculated 
self-consistently the GDE 
for DA and DB white dwarfs. The author showed that this exponent depends 
not only on the temperature 
but also on the opacity. In our work we did not calculate GDE. To show how 
different parameters 
influence the emergent spectrum, we assume values of the GDE and other 
parameters. In the Fig. \ref{teff}, we presented the distribution of the 
effective temperature over the surface of the oblate star for various GDEs. We 
assume the dimensionless angular velocity $\bar{\Omega}^2=0.60$ and the 
ratio of the mass to the equatorial radius of the neutron star $x=0.195$ and 
$\beta=0.10$, $0.15$, and $0.25$. 

\subsection{ATM24}

We calculate intensity spectra with the ATM24 code. Our model was 
described in several papers, e.g., \cite{madej91, madej04, majczyna05b}. The 
accuracy of our code was recently
improved \citep{madej17,vincent17}.

Our code solves the radiative transfer equation in the plane-parallel geometry. 
In hot atmospheres, Compton scattering plays a crucial role in forming the 
emerging spectrum and therefore should 
be treated very carefully. We investigated the scattering of X-ray photons 
with energies approaching the electron rest mass energy on free, thermal, 
relativistic electrons. We allow for any large energy 
exchange between photons and electrons during a single scattering. We assume 
the equation of state of 
ideal gas in local thermodynamical equilibrium (LTE). Note that the Compton 
scattering provides non-LTE effects in the model atmospheres. When calculating 
opacities, we
include energy-dependent free-free and bound-free opacities form all elements 
with Z$\leq$30
(atomic number) and bound-bound opacities from hydrogen, helium, and iron.

The equation of transfer was adopted from 
\citet[][see also\citealp{sampson59}]{pomraning73} and 
has the following form:
\begin{eqnarray}
&& \mu\,\frac{d I_\nu}{d \tau_\nu} = 
I_\nu-{\frac{k_\nu}{k_\nu+\sigma_\nu}}B_\nu-\left(1-{\frac{k_\nu}{k_\nu+\sigma_\nu}}
\right)J_\nu \nonumber \\
&+&
\left(1-{\frac{k_\nu}{k_\nu+\sigma_\nu}}\right) J_\nu \int\limits_0^\infty\Phi(\nu,\nu') \left(1+
{\frac{c^2}{2h{\nu'}^3}}J_{\nu'}\right) d\nu'+ \nonumber \\
&- &{\frac{k_\nu}{k_\nu+\sigma_\nu}}\left(1+{\frac{c^2}{2h\nu^3}}J_\nu\right) \nonumber \\
& \times &
\int\limits_0^\infty\Phi(\nu,\nu')J_{\nu'} {\left({\frac{\nu}{\nu'}}\right)}^3\exp\left[-{\frac{
h(\nu-\nu')}{{\rm k}T}}\right] d \nu',
\end{eqnarray}
where $k_\nu$ and $\sigma_\nu$ denote coefficients of absorption and electron 
scattering, respectively. $I_\nu$ is the energy-dependent specific intensity, 
$J_\nu$ is the mean intensity of 
radiation, and $z$ is the geometrical depth in the atmosphere. We formulated 
the exact angle-averaged
redistribution function $\Phi(\nu,\nu')$ and the Compton scattering 
cross section 
$\sigma(\nu\to\nu',\vec{n}\cdot \vec{n'})$ (for details, see \cite{madej17, 
nagirner-poutanen93b, 
nagirner-poutanen94}). The equation of transfer is solved together with the hydrostatic and 
radiative equilibrium equations because we assume that the atmosphere is 
static and that only photons transport the energy accumulated in the core of 
the neutron star. We neglect the influence of
the
magnetic field, accretion, and the effect of electron degeneracy, which is 
unimportant in the hot atmospheres. The limb-darkening/-brightening effects are 
fully taken into account in our computations.

\begin{figure}[!h]
\includegraphics[scale=0.40]{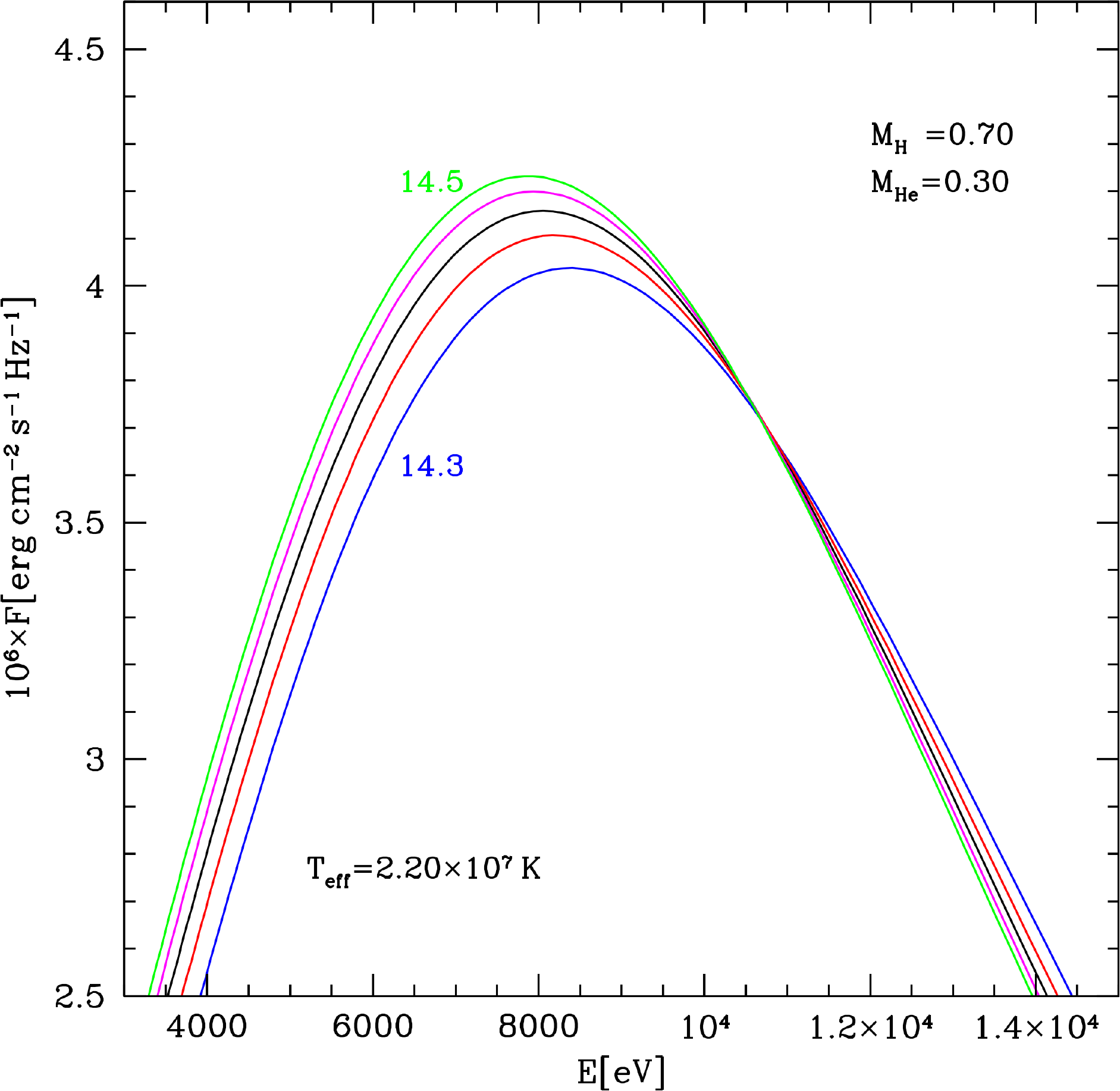}
\includegraphics[scale=0.40]{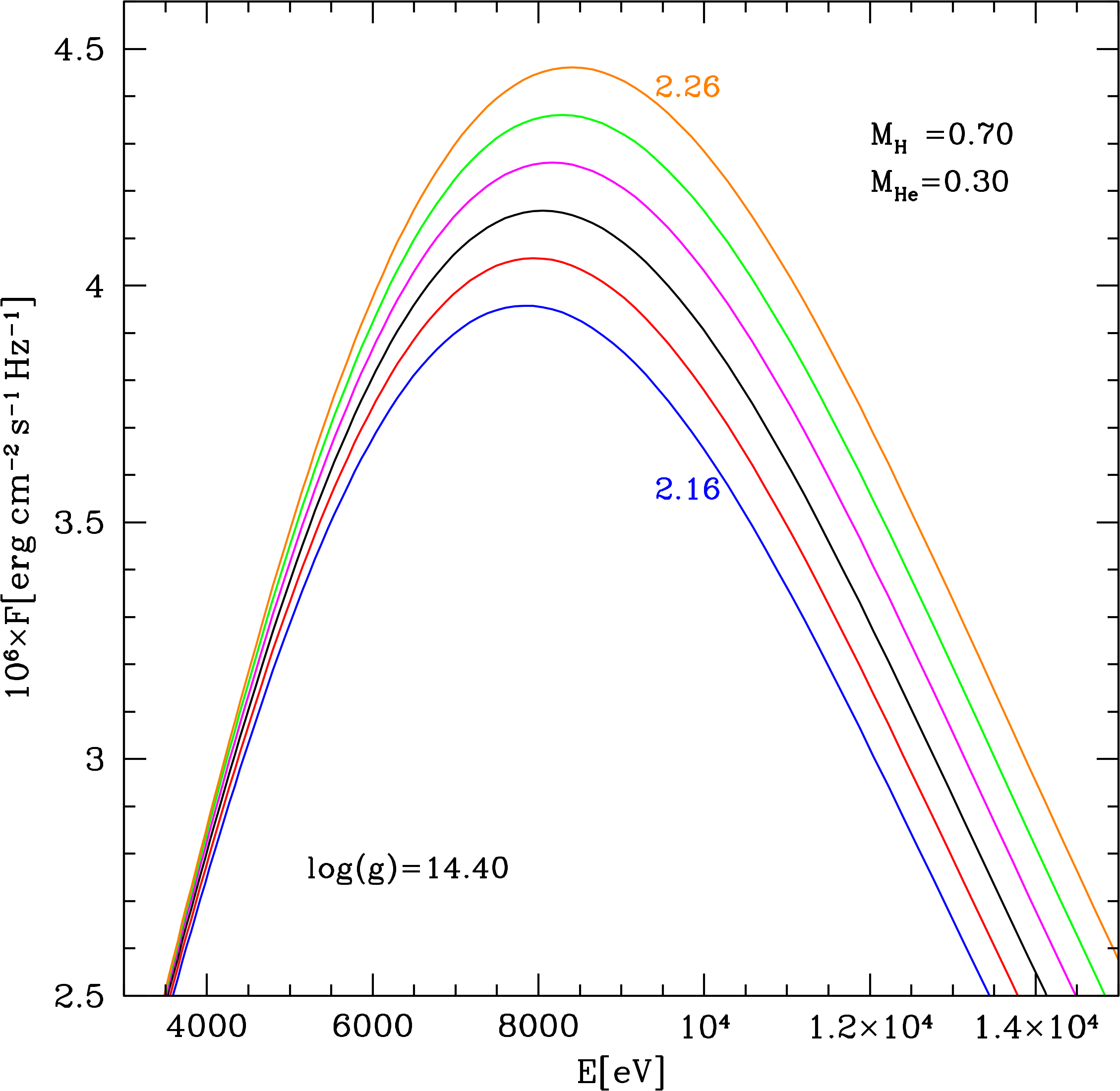}
\caption{Theoretical spectra of the unit surface calculated by the ATM24 
code for
different values of the effective
temperatures and surface gravities. In the left panel, we show the theoretical 
spectra for the effective 
temperatures $T_{\rm eff}=2.20\times 10^7\,$K and various surface gravities from
$\log(g)=14.30\,$(cgs) to $\log(g)=14.50\,$(cgs). In the right 
panel we
show spectra for the fixed surface gravity $\log(g)=14.40\,$(cgs) and various effective
temperatures from $T_{\rm eff}=2.16\times 10^7\,$K to $T_{\rm
eff}=2.26\times 10^7\,$K.}
\label{atm-widmo}
\end{figure}

We assumed that the rotating neutron star is seen at some inclination to the 
equatorial plane. 
Input parameters of the ATM24 code are the effective temperature, surface 
gravity, and the chemical 
composition. We assumed that the reference nonrotating  star has the effective 
temperature 
$T_{\rm eff}=2.20\times 10^7\,$K and the surface gravity $\log(g)=14.40\,$(cgs). The
atmosphere consists of hydrogen and helium with the cosmic mass abundance $M_{\rm H}=0.70$ and
$M_{\rm He}=0.30$.

In X-ray bursters, especially during the burst and soon after, we expect high 
temperatures of the gas, 
so $T_{\rm eff}=2.20\times 10^7\,$K is a reasonable assumption. Values of the 
effective temperature 
and the logarithm of the surface gravity of the spherical star were taken 
arbitrarily. The chemical 
composition of X-ray burst sources is very poorly constrained. It is frequently assumed that the 
atmosphere consists of hydrogen and helium in arbitrarily chosen proportions. 
For example, \cite{galloway_ea06} investigated Eddington-limited bursts from 
4U~1636-536. For this source, a hydrogen-rich atmosphere with $M_{\rm 
H}=0.70$ as well as a pure helium atmosphere were
investigated. Therefore, we assume chemical composition with cosmic hydrogen 
abundance. 

The rotating star is oblate, therefore the gravity and the effective 
temperature are different at 
various latitude angles. We calculated the local effective gravity and the 
local effective temperature 
according to the formulae given by Equations \ref{eq:geff} and \ref{eq:teff}. 
These local values of the 
gravity and the effective temperature are input parameters for the calculation 
of monochromatic specific intensities at various latitude angles.
These intensities were next integrated over the whole surface to obtain the spectrum as seen 
by a distant observer. 

For a given eccentricity of the distorted star, we were able to compute the 
distance from the 
center of gravity for each patch on the surface and its area (in cm$^2$), its 
latitude, and the 
azimuth as seen by the observer. Simultaneously, we determined the cosine of 
the projection angle as 
seen by the observer. We determined the effective gravity as a function 
of the latitude.
Therefore,
we know all variables that are necessary to integrate the emergent intensity 
over various patches to 
obtain the final monochromatic luminosity of the distorted star.

In our ATM24 code, we include the limb-darkening effect and now we add the 
gravitational darkening 
effect. An example of flux spectra of the spherical star calculated by 
the ATM24 code is presented in Figure \ref{atm-widmo}. In the left panel, we 
compare the spectra for the same chemical compositions 
given above and the same effective temperatures but different surface gravities 
from $\log(g)=14.30$ 
up to 14.50~(cgs). With increasing surface gravity, spectra become 
softer, and their maximum
shifts
toward lower energies. The right panel shows our model spectra calculated for the same surface 
gravity and different effective temperatures from $T_{\rm eff}=2.16\times 10^7$ up to $2.24\times
10^7\,$K. With increasing temperature, the maximum of the spectra shifts toward 
higher energies.

\section{Spectra of the rotating neutron star}
We computed a small grid of theoretical spectra of the rotating neutron stars 
(56 model spectra). 
Because the calculations are very time consuming, we divided the star into 
18x36 points in latitude and 
azimuthal angles. As was mentioned above, we assumed parameters of the rotating star like the 
effective temperature of the spherical star $T_{\rm eff}=2.20\times 10^7\,$K, surface gravity of the 
nonrotating star $\log(g)=14.40\,$(cgs), chemical composition ($M_{\rm H}=0.70$ 
and $M_{\rm 
He}=0.30$)  and mass to radius ratio $x=0.195$. We computed our theoretical spectra for different 
values of the other parameters: the dimensionless angular velocity $\bar{\Omega}^2=0.30$ and $0.60$, 
and few inclination angles from $i=0^\circ$ up to $90^\circ$ with step $\Delta i=10^\circ$. We note 
that the both assumed values of the dimensionless angular velocity are relevant 
for the fast-rotation limit.

\begin{figure}[!h]
\begin{center}
\includegraphics[scale=0.45]{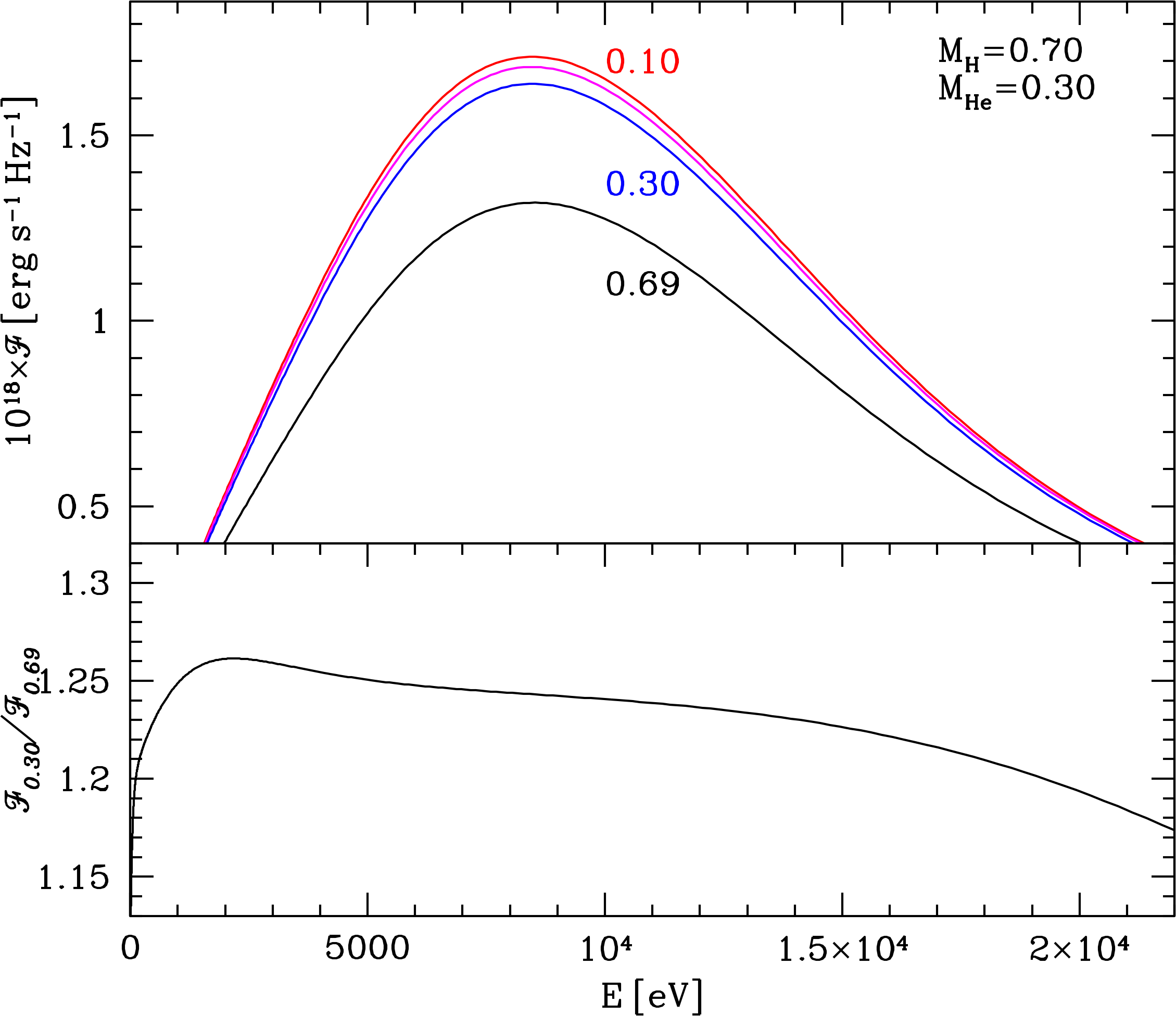}
\end{center}
\caption{The top panel shows the theoretical spectra of the flattened neutron 
star for a given parameters: 
$T_{\rm eff}=2.20\times 10^7\,$K, $\log(g)=14.40\,$(cgs), $i=30^\circ$, 
$\bar{\Omega}^2=0.60$ 
and different values of the eccentricity of the star $e=0.10$ (red line), 0.20 
(magenta), 0.30 (blue), and 0.69 (black). The lower panel shows the ratio of 
the monochromatic flux ($\mathcal{F}_{0.30}$) 
emitted by the ellipsoidal star with $e=0.30$ and flux ($\mathcal{F}_{0.69}$) 
emitted by a star with 
$e=0.69$.}
\label{ecc}
\end{figure}

The eccentricity of the rotating neutron star depends on the angular velocity 
of the rotation. The formula that connects these two variables depends on the 
equation of state of the matter building up 
neutron star. Such approximate formulae were proposed by various authors 
\cite[see e.g.,][]{algendy14,silva_ea21}. In our paper, we used the relation 
from \cite{silva_ea21}.
\begin{equation}
e=0.251+0.935\bar{\Omega}^2+0.709x+0.030\bar{\Omega}^2x-0.472\bar{\Omega}^4-2.427x^2
\end{equation}
Here, we quoted Eq.19 from \cite{silva_ea21} using variables defined in our 
paper. For the assumed 
$x=0.195$ and $\bar{\Omega}^2=0.30$ and $0.60$, the eccentricity of the 
flattened star is equal to
$e=0.54$ and 0.69, respectively. These values of the eccentricity were used 
during the calculations of 
theoretical spectra of rotating neutron star. In some cases where we show 
the influence of the shape 
of the star on the emergent spectrum, we thread the eccentricity as a free 
parameter, independent of the angular velocity (see Fig. \ref{ecc}).

Assuming $\bar{\Omega}^2=0.30$, and 0.60, $x=0.195$ and two representative 
equations of state 
\citep{algendy14} calculated the frequency of the neutron star rotation (see 
Tables 2 and 3 in their paper). For both dimensionless angular velocities they 
obtained high frequencies, e.g., for 
$\bar{\Omega}^2=0.30$, $M=1.80\,M_\odot$ and $R=13.6\,$km (EOS HLPS2) 
the frequency is equal to 
$\nu=842\,$Hz whereas for $\bar{\Omega}^2=0.60$, $M=1.99\,M_\odot$ and 
$R=15.1\,$km (EOS HLPS2), $\nu=1084\,$Hz. The highest measured frequency of the 
rotation is equal to 716 Hz in the case of pulsar PSR~J1748-2446ad 
\citep{hessels_ea06}. 
For X-ray bursters, the fastest spin of the neutron star equals to 
$\nu=620\,$Hz for 4U~1608-522 
\citep{muno_ea02}. The maximal value of the frequency of the neutron star 
rotation depends on the 
equation of state, and it is as high as 1200~Hz \cite[see e.g.,][]{haensel07}.

Figure \ref{ecc} (top panel) shows theoretical spectra of the rotating star 
with different 
eccentricities for the same inclination angle $i=30^\circ$. We note that these 
spectra have very similar shapes. It may appear that these spectra differ only 
in the normalization. For this 
reason, in the lower panel of the Figure \ref{ecc} we show the ratio of the 
monochromatic flux emitted 
by the star with $e=0.30$ and the flux of the star with $e=0.69$. If these 
spectra differ only by the 
normalization, this ratio should be constant across the whole range of 
photon energies, which is not the case presented in the figure.
Of course, for lower differences of the eccentricity, the ratio is closer to 
constant so the spectra have 
more similar shapes. It is especially important for methods of mass and radius 
determination that used 
the distance to the source \citep[see e.g., the review by ][]{sudip10}. Our 
mass and radius determination 
method \citep[e.g.,][]{majczyna05} is independent of the distance; therefore 
value of the obtained 
normalization does not affect our results. 

\begin{figure}[!h]
\includegraphics[scale=0.4]{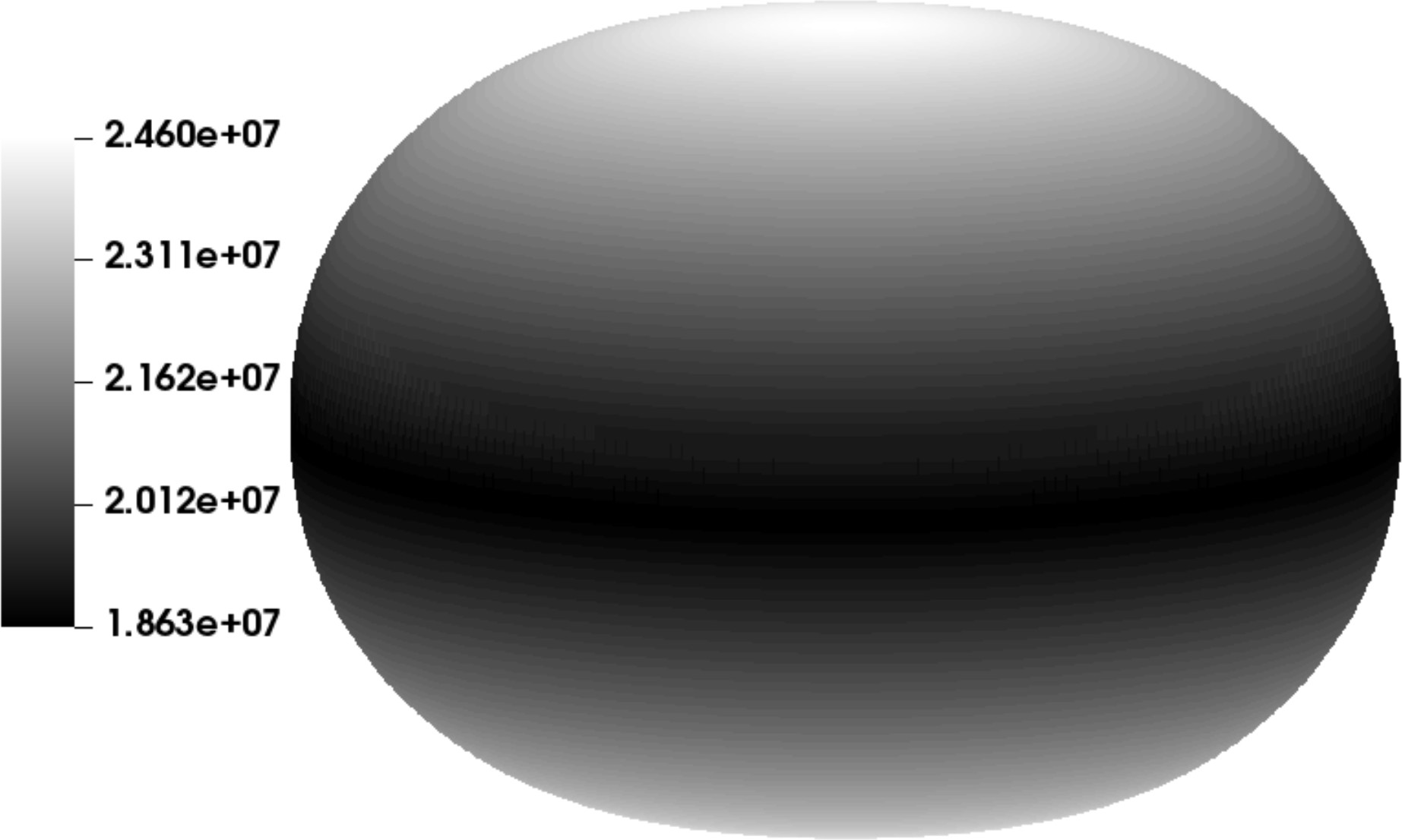}
\includegraphics[scale=0.4]{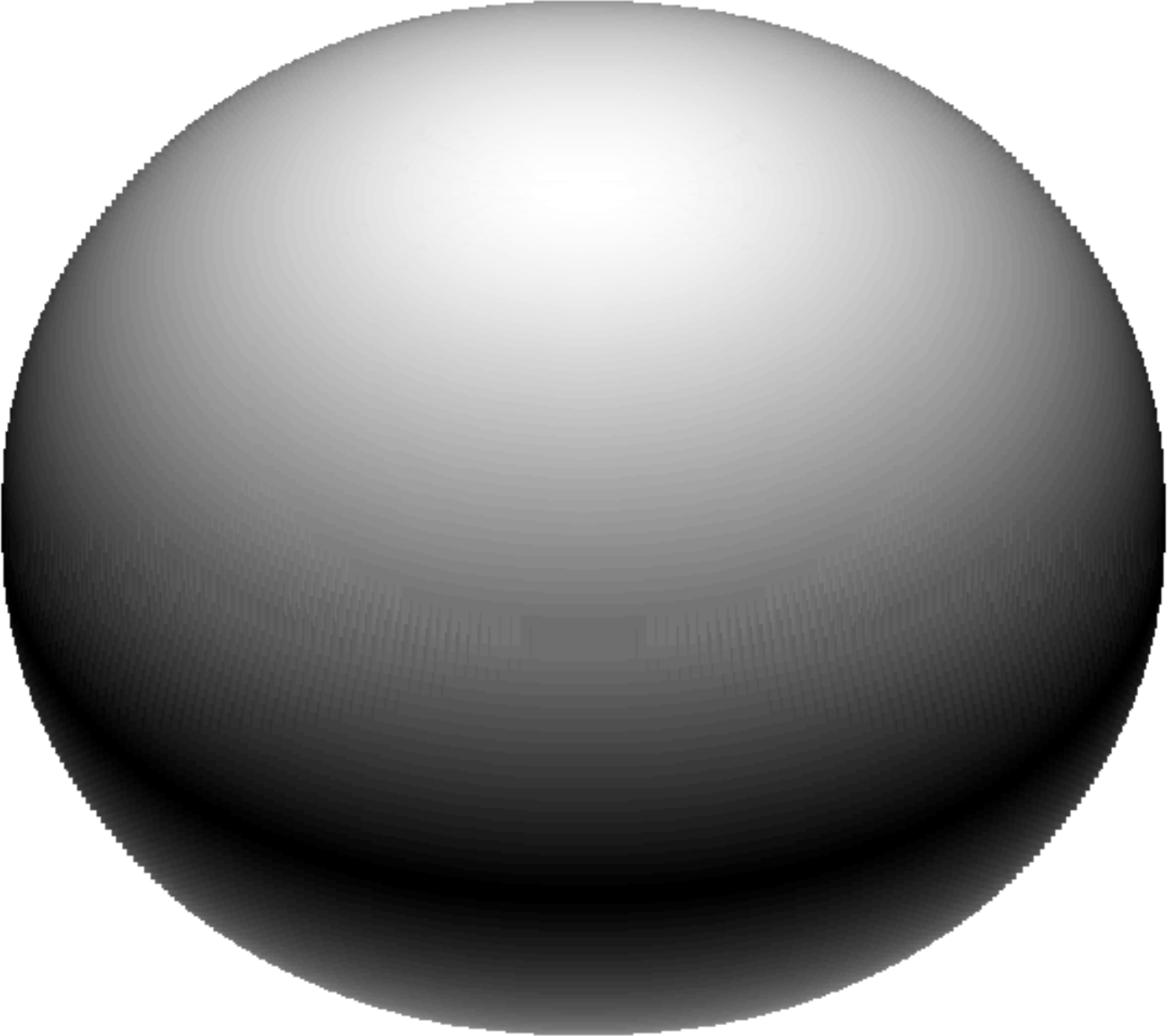}
\includegraphics[scale=0.3]{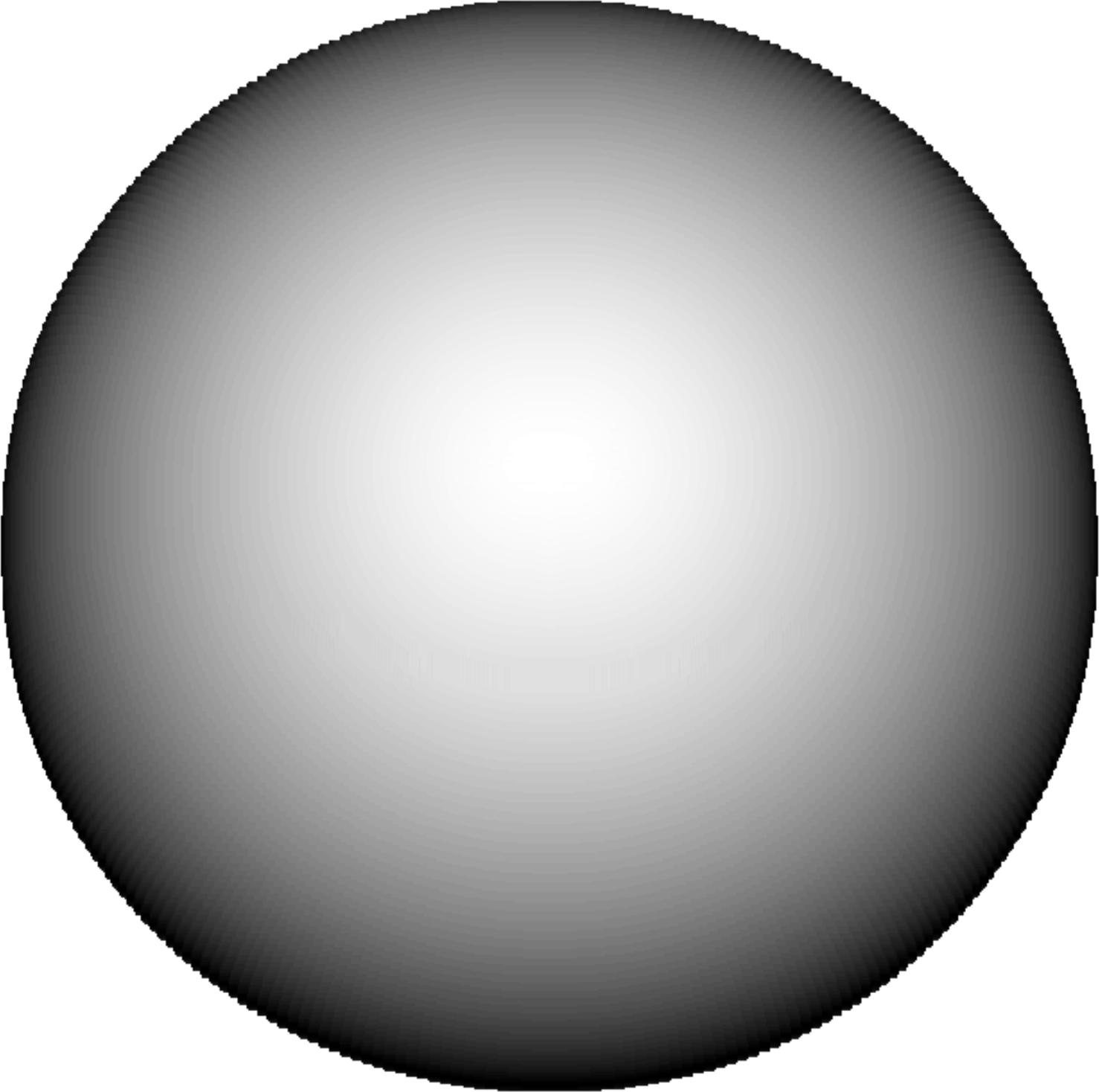}
\caption{Distribution of the local effective temperature over the surface of the neutron star seen 
by a distant observer. We assume parameters of our reference model for $T_{\rm eff}$ and 
$\log(g)$ and the gravitational darkening exponent $\beta=0.25$, dimensionless angular velocity 
$\bar{\Omega}^2=0.60$ and different inclination angles $i=10^\circ$ (left panel), $i=40^\circ$ 
(middle panel), and $i=80^\circ$ (right panel).}
\label{elipsoida-upth}
\end{figure}

Figure \ref{elipsoida-upth} shows the temperature distribution on the star 
seen at different inclination 
angles $i=30^\circ$ (left panel), $i=60^\circ$ (middle), and $i=80^\circ$ 
(right panel). The last panel of this figure presents the star seen almost at 
the pole on, where the tem\-pe\-ra\-tu\-re has 
the highest value. 

\begin{figure}[!h]
\begin{center}
\includegraphics[scale=0.38]{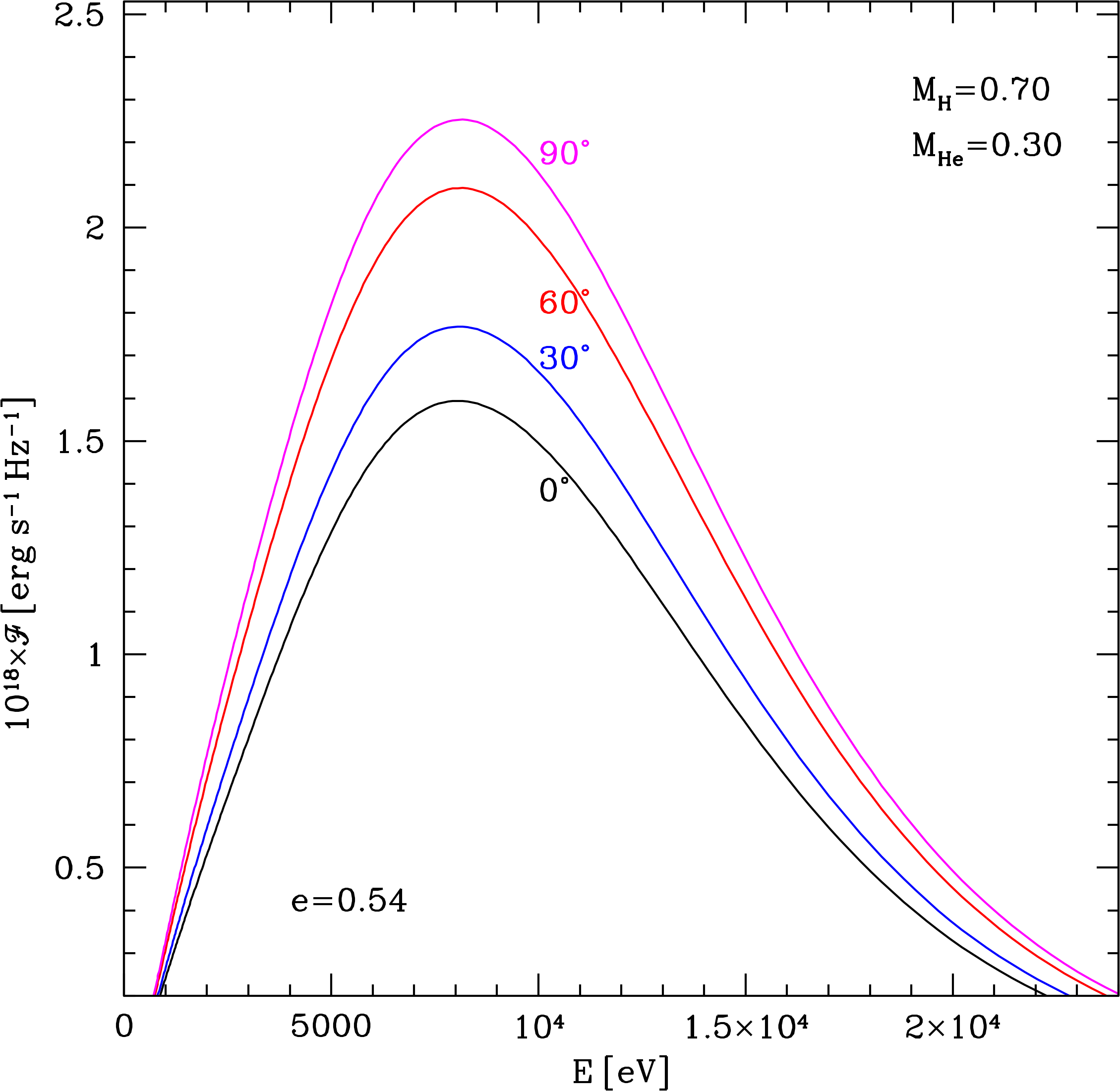}
\includegraphics[scale=0.38]{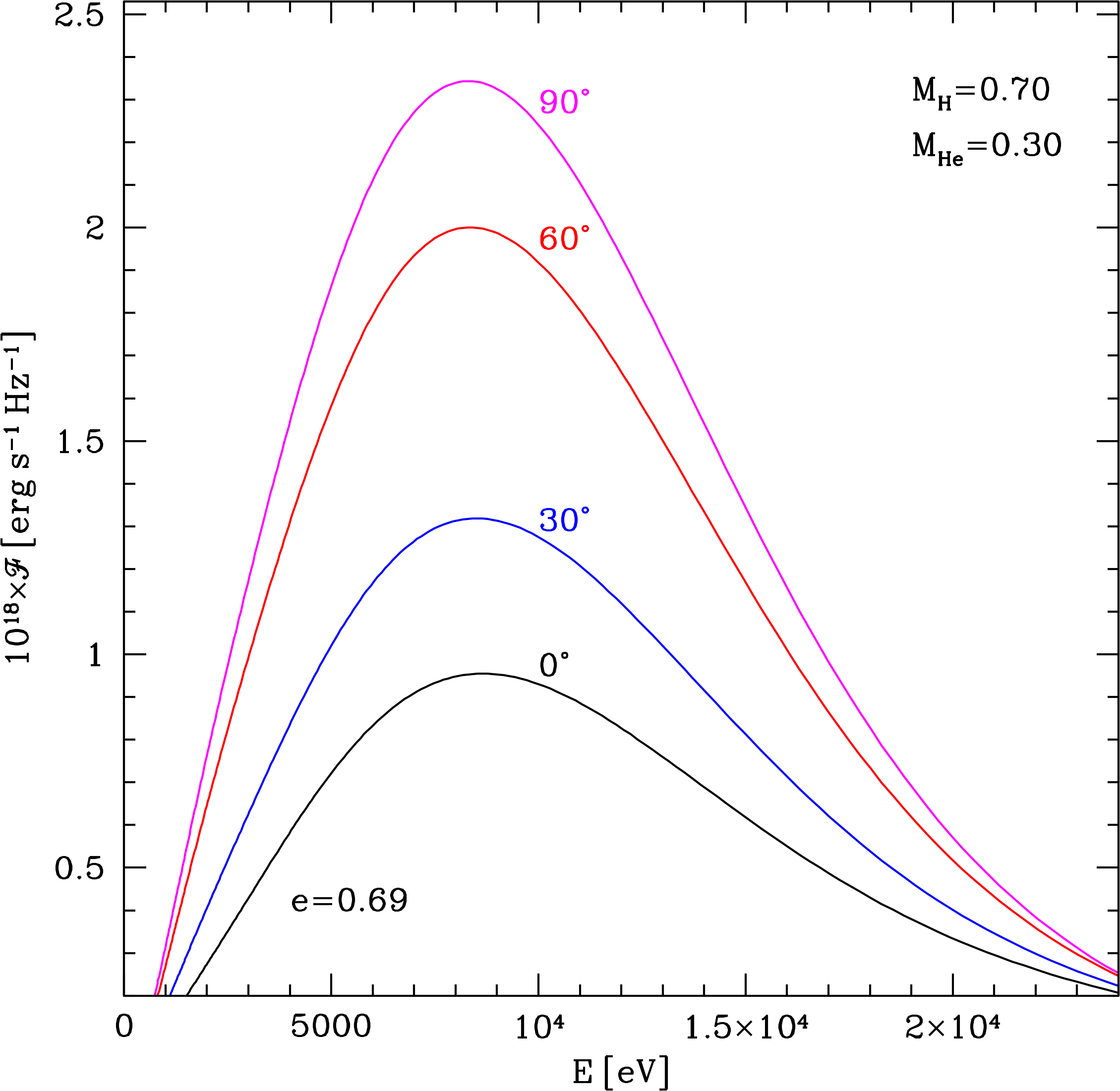}
\end{center}
\caption{Theoretical spectra of the rotating neutron star for various inclination angles. We assumed 
parameters $T_{\rm eff}=2.20\times 10^7\,$K, $\log(g)=14.40\,$(cgs), 
$\beta=0.25$, and two 
values of the dimensionless angular velocity $\bar{\Omega}^2=0.30$ (left panel) and 
$\bar{\Omega}^2=0.60$ (right 
panel).}
\label{upth}
\end{figure}

The spectral shape significantly depends on the inclination angle as was shown 
in the Figure 
\ref{upth}. An inclination angle equal to $i=90^\circ$ denotes that an 
observer sees the star at the pole, 
whereas $i=0^\circ$ is in the equatorial plane. As the inclination angle 
increases the observer 
sees hotter and hotter areas, but the spectrum does not become harder and its maximum shifts toward 
lower energies. This slightly unintuitive effect is connected with 
the different contributions of 
spots with different temperatures when the inclination angle changes. Hotter areas located at the 
poles are smaller than those colder ones located at the equatorial plane.

\begin{figure}[!h]
\begin{center}
\includegraphics[scale=0.38]{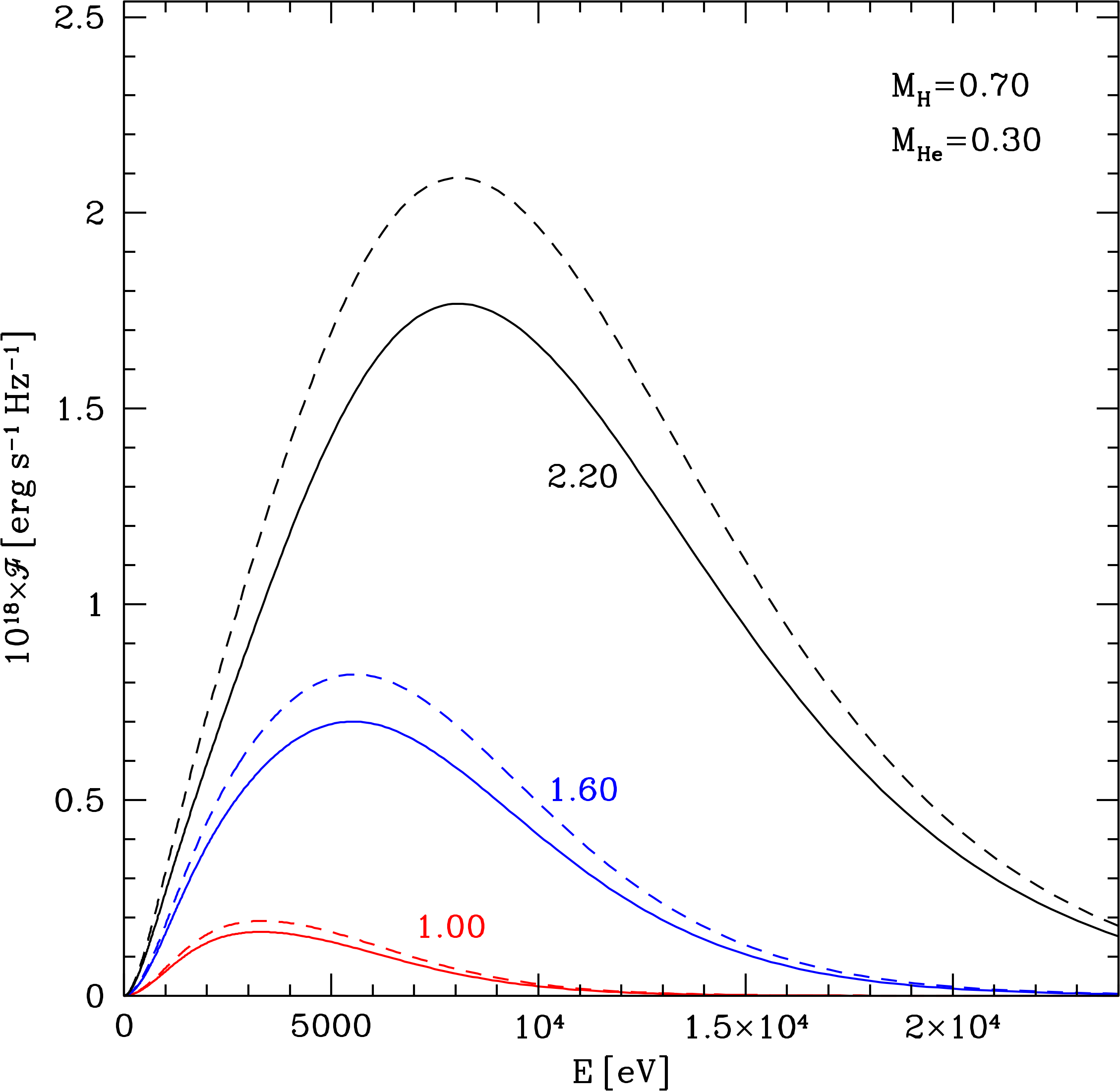}
\includegraphics[scale=0.38]{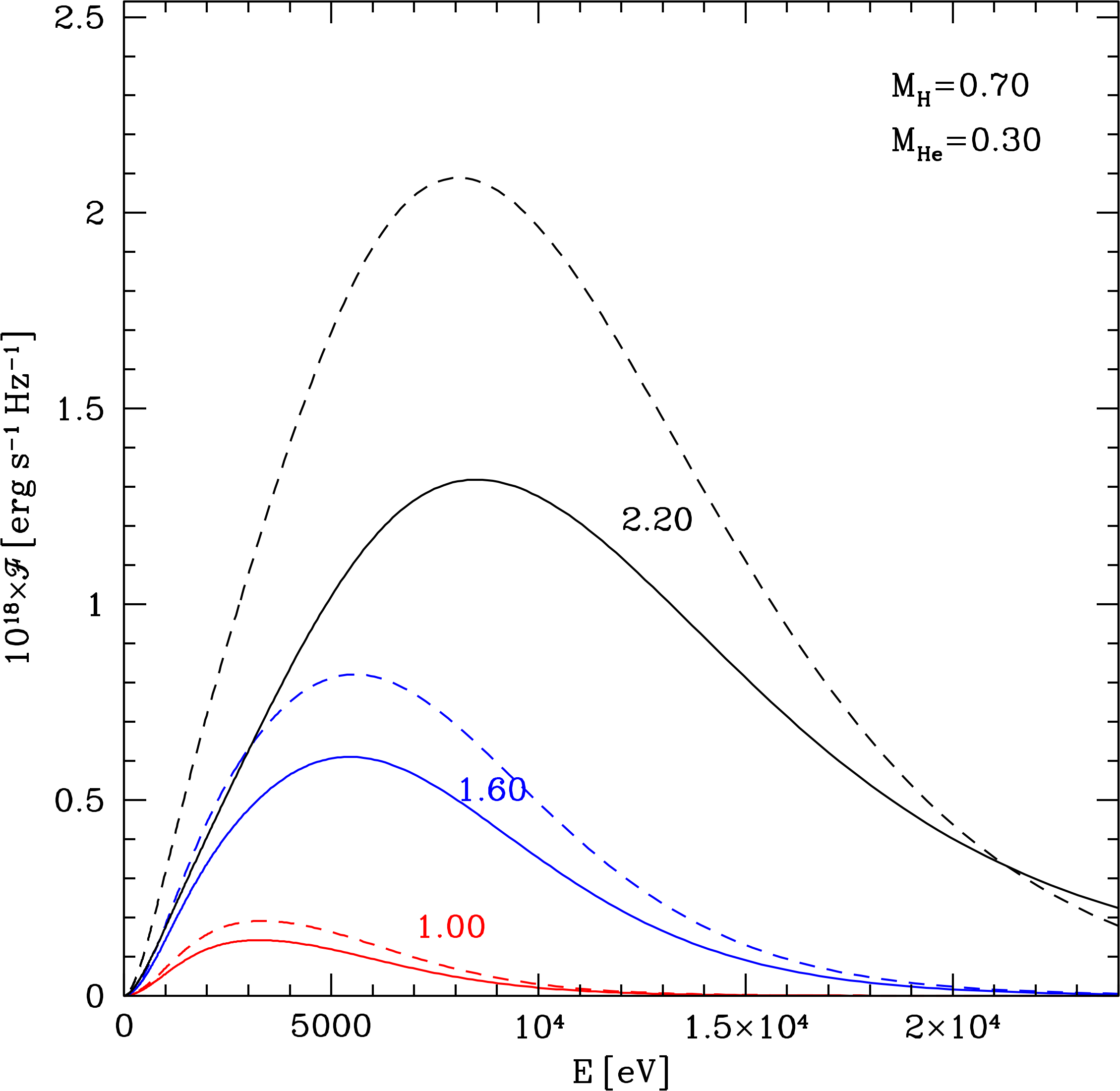}
\end{center}
\caption{Theoretical spectra of the rotating star for different effective temperatures of 
the spherical neutron star. We fixed following the parameters: 
$\log(g)=14.40$, $\beta=0.25$, 
$i=30^\circ$, and $\bar{\Omega}^2=0.30$; $e=0.54$ (left panel) and 
$\bar{\Omega}^2=0.60$; 
$e=0.69$ (right panel). The black solid curve denotes spectrum of the oblate star calculated 
for assumed temperatures of the spherical star $T_{\rm eff}=2.20\times 10^7\,$K 
whereas blue ones assume $T_{\rm eff}=1.60\times 10^7\,$K and red $T_{\rm 
eff}=1.00\times 10^7\,$K. Black, blue, and red 
dashed curves denote spectra of the spherical stars with appropriate 
temperatures.}
\label{darkteff}
\end{figure}

Figure \ref{darkteff} shows spectra emitted by the rotating star calculated for various effective 
temperatures of the spherical star. The left panel shows spectra of the 
neutron star rotating with 
$\bar{\Omega}^2=0.30$, whereas the right one is rotating with 
$\bar{\Omega}^2=0.60$. In both cases, we assume 
the logarithm of gravity of the nonrotating star 
$\log(g)=14.40\,$(cgs). The dashed line denotes 
spectra of the spherical star. This star has various values of temperature ($T_{\rm eff}=1.0\times 
10^7\,$K, $1.6\times 10^7\,$K and $2.2\times 10^7\,$K respectively) constant 
over the surface. The maxima of spectra with gravitational darkening effect are 
shifted toward higher energies, and spectra are harder than these emitted by 
the spherical star.

\begin{table}[!h]
\caption{Bolometric luminosities and effective temperatures of the rotating, flattened neutron 
star with two dimensionless angular velocities $\bar{\Omega}^2$ and a few 
inclination 
angles $i$. We assumed the effective temperature of the spherical star $T_{\rm sph}=2.20\times 
10^7\,$K. The bolometric luminosity of the spherical star is equal to $L_{\rm sph}=8.179\times 
10^{37}\,{\rm erg}\,{\rm s}^{-1}$ and color temperature $T_c=3.31\times 10^7\,$K. We assume that the 
equatorial 
radius of the oblate star and the radius of the spherical star is equal to $R=7\,$km.}
\label{lbol}
\begin{center}
\begin{tabular}{|l|c|c|c|c|c|c|}
\multicolumn7c{$\bar{\Omega}^2=0.30$ ($e=0.54$)} \\ \hline
Model & $L_{\rm bol}\times 10^{37}$ & $L_{\rm bol}/L_{\rm sph}$ & $T_{\rm eff}\times 
10^{7}\,$K & $T_{\rm eff}/T_{\rm sph}$ & $T_c\times 10^7\,$K & $f_{\rm c}$ \\ \hline
%
$i=0.0$ & 6.365 & 0.78 & 2.1165 & 0.962 & 3.31199 & 1.565 \\
$i=30$  & 7.103 & 0.87 & 2.1753 & 0.989 & 3.31199 & 1.523 \\
$i=90$  & 9.159 & 1.12 & 2.3181 & 1.054 & 3.36002 & 1.449 \\
\hline
\multicolumn7c{$\bar{\Omega}^2=0.60$ ($e=0.69$)} \\ \hline
Model & $L_{\rm bol}\times 10^{37}$ & $L_{\rm bol}/L_{\rm sph}$ & $T_{\rm eff}\times 
10^{7}\,$K & $T_{\rm eff}/T_{\rm sph}$ & $T_c\times 10^7\,$K & $f_{\rm c}$ \\ \hline
%
$i=0.0$ & 4.805 & 0.59 & 2.0007 & 0.909 & 3.55911 & 1.779 \\
$i=30$  & 6.164 & 0.75 & 2.1292 & 0.968 & 3.50826 & 1.648 \\
$i=90$  & 9.827 & 1.20 & 2.3926 & 1.088 & 3.40870 & 1.425 \\
\hline
\end{tabular}
\end{center}
\end{table}

Table \ref{lbol} presents examples of a few values of the effective and color 
temperatures for 
models with two values of the dimensionless angular velocities $\bar{\Omega}^2=0.30$ and 
$0.60$ and a few inclination angles $i$. As to be expected, stars seen in the 
equatorial plane 
($i=0^\circ$) have the lowest effective temperatures ($T_{\rm eff}=2.117\times 10^7\,$K and $T_{\rm 
eff}=2.001\times 10^7\,$K for $\bar{\Omega}^2=0.30$ and $0.60$, respectively), 
the lowest color 
temperatures, and the lowest bolometric fluxes. On the contrary, stars seen 
pole on are brighter and 
apparently hotter. The last column of the Table \ref{lbol} contains spectral 
hardening factor 
$f_c=T_c/T_{\rm eff}$ (here, $T_c$ is the color temperature whereas $T_{\rm 
eff}$ is the effective 
temperature). The color temperature is calculated from the Wien displacement law $h\nu_{\rm 
max}=2.82kT$. The value of the hardening factor depends on the 
inclination angle as well as on the dimensionless angular velocity. The 
largest values of $f_{\rm c}$ were obtained for the star seen pole on 
whereas the lowest ones were for the star seen in the equatorial plane. For 
the spherical star with reference parameters $T_{\rm eff}=2.20\times 10^7\,$K 
and $\log(g)=14.40\,$(cgs), the spectral hardening factor is equal to 
$f_{\rm c}=1.50$ and differs from values for the flattened star given in 
the last column of the table.

\begin{figure}[!h]
\begin{center}
\includegraphics[scale=0.40]{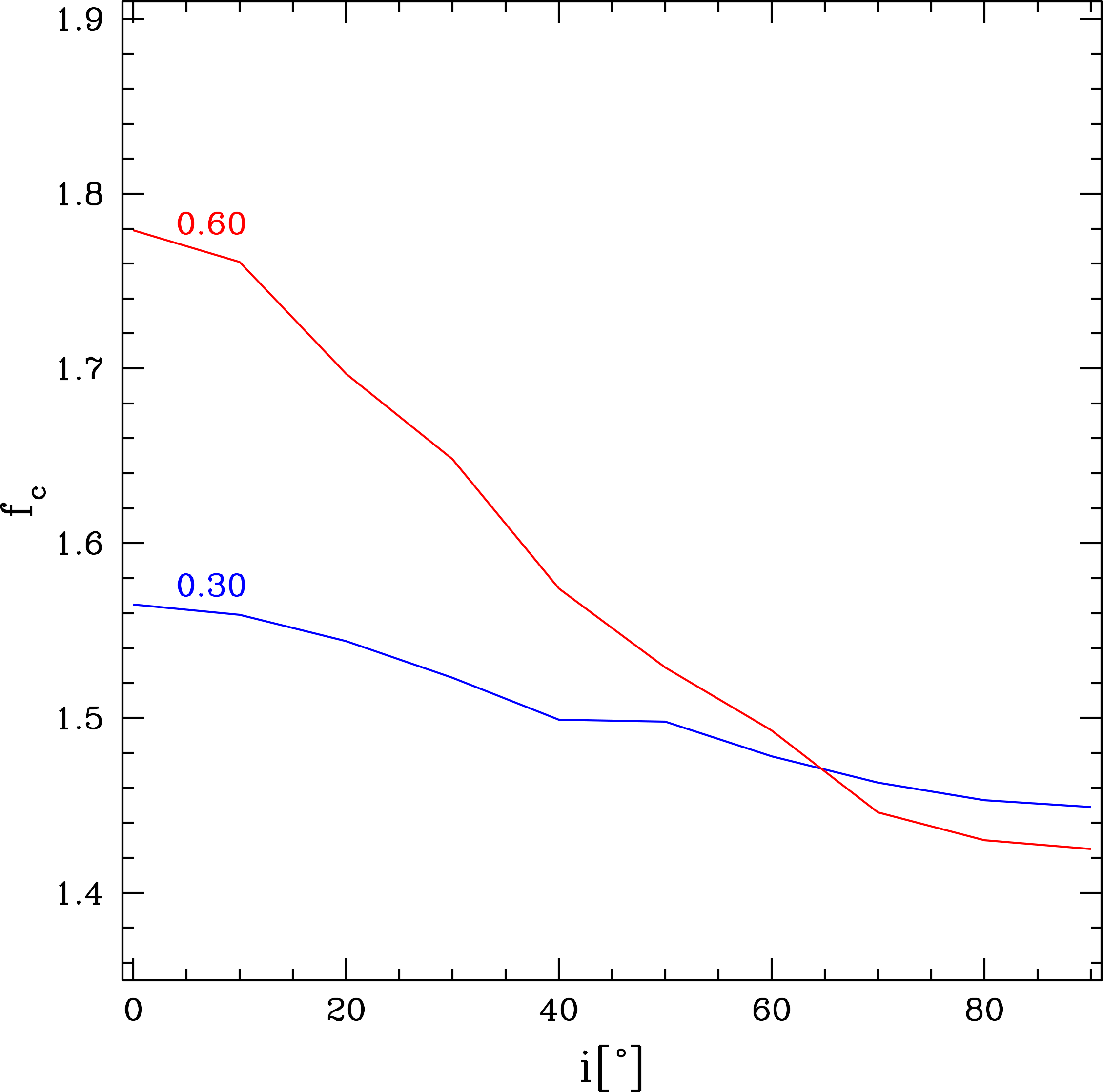}
\end{center}
\caption{Spectral hardening factor $f_c$ for a flattened star with our 
reference parameters seen 
at different inclination angles $i$ and two sets of parameters 
$\bar{\Omega}^2=0.30$ and $e=0.54$ (blue line labeled 0.30) and 
$\bar{\Omega}^2=0.60$ and $e=0.69$ (red line labeled 0.60).}
\label{fc}
\end{figure}

Figure \ref{fc} presents the dependence of the spectral hardening factor for 
a rotationally flattened neutron star 
with two values of the dimensionless angular velocity $\bar{\Omega}^2=0.30$ and 
$0.60$ on inclination angles from $i=0^\circ$ (star seen in the equatorial 
plane) up to $i=90^\circ$ 
(seen pole on). For stars with higher values of the angular velocity of the 
rotation and therefore an 
eccentricity, the spectral hardening factor has larger values and changes more rapidly with the 
inclination angle than in the case of the stars with a lower value of 
$\bar{\Omega}^2$.

\section{Summary and discussion}

In our paper, we presented spectra of a rotationally distorted neutron star 
considering also the 
case of fast rotation. We have shown the substantial influence of both the 
dimensionless angular 
velocity $\bar{\Omega}^2$ and the inclination angle $i$ on the emitted spectra. 

Using the ATM24 code we calculated a small sample grid of theoretical 
spectra integrated 
over the distorted surface of a sample rotating neutron star seen by a distant 
observer at various 
inclination angles. We assumed the following parameters: two dimensionless 
angular velocities $\bar{\Omega}^2=0.30$ 
and $0.60$, the effective temperature of the nonrotating star $T_{\rm 
eff}=2.20\times 10^7\,$K, the logarithm of the surface gravity of the 
spherical star $\log(g)=14.40\,$(cgs) and inclination angles 
from $i=0^\circ$ to $i=90^\circ$ with the step $\Delta i=10^\circ$. We assume that the atmosphere is 
a mixture of hydrogen and helium with $M_{\rm H}=0.70$ and $M_{\rm He}=0.30$.

The effective gravity and the effective temperature (determined from von 
Zeipel law) are strong functions of the dimensionless angular velocity 
$\bar{\Omega}^2$ (see Eq.\ref{eq:geff}). 
Therefore, $\bar{\Omega}^2$ belongs to parameters that play a crucial role in 
the formation of the 
emergent spectrum. The eccentricity of the oblate star depends on 
$\bar{\Omega}^2$ and was 
obtained by the approximate (independent of the equation of state) formula 
from \cite{silva_ea21}.

\cite{suleimanov_ea20} have shown the spectra of the rotating neutron star as 
a function of the 
inclination angle and the angular velocity of the star. They applied a diluted 
blackbody spectrum to 
describe the local emission corrected by the hardening factor. The authors 
assumed slow-rotation 
approximation during the calculation of the effective gravity. Their Constant 
Relative Flux model is the 
same as our model of a flattened neutron star with the effective temperature 
proportional to the 
local effective gravity as $T_{\rm eff}(\theta)\sim g(\theta)^{0.25}$. Therefore, our model 
corresponds to the upper panel of the Figure~6 in their paper. The lower panel 
of Fig.6 in their paper 
simply ignores the von Zeipel law. The paper mentioned above was based on the 
code defined in 
\cite{suleimanov12}, where the equation of transfer has part of 
the stimulated Compton scattering terms artificially transferred to the 
denominator (see Eq. 12 their 
paper). As a result expression for the source function was depleted by one of 
the Compton-induced 
scattering terms (see Eq. 14). The impact of such an arbitrary change in their 
paper on the atmosphere 
models near the Eddington limit is unknown. In the ATM 24 code, such an 
artificial manipulation does not 
exists.

In our model we applied fast rotation approximation; therefore, we included 
the quadrupole moment of the mass distribution in the neutron star interior. 
We calculated very carefully specific intensities 
$I_\nu(T_{\rm eff}(\theta),g_{\rm eff}(\theta))$ at each patch on the surface. These specific 
intensities are next integrated over the surface of the neutron star to obtain 
spectra that could be 
seen by the distant observer \citep{agata17,agata18}. Spectra presented 
in this paper are not
corrected for the relativistic 
effects, which will be the subject of our future work. In our paper, we did 
not include effects of the 
special relativity (relativistic Doppler broadening), which change the shape 
of the spectrum. This 
has been investigated by \cite{baubock_ea15} for a perfect black body spectrum. Authors concluded 
that changes due to Doppler broadening are small compared to the gravitational redshift.

In this paper, we clearly showed that the gravitational darkening effect should 
be included in 
realistic calculations of the rotating neutron star spectra. This effect 
modifies the shape of the 
emerging spectrum and therefore affects the determination of neutron star 
parameters obtained by 
spectral fitting of real data. 

\acknowledgements
This work was supported by grant No. 2015/17/B/ST9/03422 from the Polish National Science Center.

\bibliographystyle{aa} 
\bibliography{ns1}

\end{document}